\documentclass[floatfix,nofootinbib,twocolumn,showpacs,superscriptaddress, preprintnumbers,amssymb,letterpaper,prl]{revtex4-2}

\usepackage{lineno}
\usepackage{textgreek}

\usepackage{dcolumn}
\usepackage{bm}
\usepackage{amsmath}
\usepackage{multirow}
\usepackage{graphicx}
\usepackage{float}
\usepackage[colorlinks=true,allcolors=blue]{hyperref}

\usepackage[encapsulated]{CJK}
\usepackage[english]{babel}
\usepackage{siunitx}
\usepackage{comment}
\usepackage{textcomp}

\begin{document}


 \newcommand{\re}{\mathop{\mathrm{Re}}}
 \newcommand{\im}{\mathop{\mathrm{Im}}}
 \newcommand{\D}{\mathop{\mathrm{d}}}
 \newcommand{\E}{\mathop{\mathrm{e}}}
 \newcommand{\unite}[2]{\mbox{$#1\,{\rm #2}$}}
 \newcommand{\myvec}[1]{\mbox{$\overrightarrow{#1}$}}
 \newcommand{\mynor}[1]{\mbox{$\widehat{#1}$}}
\newcommand{\rmsemit}{\mbox{$\tilde{\varepsilon}$}}
\newcommand{\mean}[1]{\mbox{$\langle{#1}\rangle$}}
\newcommand{\warp}{{\sc Warp }}
\newcommand{\astra}{{\sc Astra }}
\newcommand{\elegant}{{\sc Elegant }}
\newcommand{\astragenerator}{{\sc AstraGenerator }}
\newcommand{\mafia}{{\sc Mafia }}
\mathchardef\mhyphen="2D
\preprint{~~~~}

\title{Direct Measurement of Eigenemittances Transfer to Projected Emittances \\
via Phase-Space Decoupling for an Electron Beam  }

\author{T. Xu}
\email{xu@niu.edu}
\affiliation{Northern Illinois Center for Accelerator \& Detector Development and Department of Physics, Northern Illinois University, DeKalb, IL 60115, USA} 
\author{S. Doran}
\affiliation{Argonne National Laboratory, Lemont, IL 60439, USA}
\author{W. Liu}
\affiliation{Argonne National Laboratory, Lemont, IL 60439, USA}
\author{P. Piot}
\affiliation{Northern Illinois Center for Accelerator \& Detector Development and Department of Physics, Northern Illinois University, DeKalb, IL 60115, USA} 
\affiliation{Argonne National Laboratory, Lemont, IL 60439, USA}
\author{J. G. Power}
\affiliation{Argonne National Laboratory, Lemont, IL 60439, USA}
\author{C. Whiteford}
\affiliation{Argonne National Laboratory, Lemont, IL 60439, USA}
\author{E. Wisniewski}
\affiliation{Argonne National Laboratory, Lemont, IL 60439, USA}

\date{\today}

\begin{abstract}
Phase-space partitioning offers an attractive path for the precise tailoring of complex dynamical systems. In Beam Physics, the proposed approach  involves ($i$) producing beams with cross-plane correlations to control kinematical invariants  known as eigenemittances and ($ii$) mapping them to invariants of motion associated with given degrees of freedom via a decoupling transformation. Here we report on the direct experimental demonstration of the mapping of eigenemittances to transverse emittances for an electron beam. Measured phase space density confirms the generation of beams with asymmetric transverse emittance ratio$>200$ consistent with the initiated eigenemittance values. The results could have broad applications to other fields where invariants are sometimes used to describe coupled classical system quantum systems with mixed states. 
\end{abstract}

\maketitle
The evolution of $N$-body systems governed by linear Hamiltonian dynamics is often described statistically using ensemble-averaged quantities. In applications where the phase-space probability density function (PDF) of such a system is of interest, averaged kinematical invariants of the motion are introduced~\cite{dragt-1992-a}. This practice is common to, e.g., quantum mechanics~\cite{bartlett-2012-a}, optics~\cite{simon-2000-a} and charged-particle beam physics~\cite{neri-1990-a}. In systems where the three degrees of freedom (DOF) are decoupled, invariants of the motion can be introduced along each DOF. For instance, consider the position-conjugate momentum pair ($q_\ell,p_\ell$), with $\ell=x,y,z$, associated with a DOF, the moment invariant $\varepsilon_\ell \equiv [\langle{q_\ell^2}\rangle \langle{p_\ell^2}\rangle-\langle{q_\ell p_\ell}\rangle^2]^{1/2}$ is often introduced as it statistically quantifies Liouville's theorem related to the conservation of phase-space density under linear transformations of the coordinates~\cite{liouville-1838-a}. In Beam Physics $\varepsilon_\ell$,  termed ``emittance", plays a critical role in quantifying the beam quality~\cite{lapostolle-1971-a}. Similar invariants are sometimes introduced in Optics with moments of the Wigner function to quantify the impact of, e.g., aberrations~\cite{dodonov-2000-a}. Likewise, in quantum systems with mixed states, generalized moment invariants help to study the time evolution of these states. In beam physics, the eigenemittances are a generalization of the projected emittances to the case of beams with coupled DOFs~\cite{duffy-2016-a}. The eigenemittances can be converted into projected emittances by symplectic transformations. In the absence of damping or particle losses, the eigenemittances determine the ultimate limits on the achievable beam brightness. The initial eigenemittances set the final beam emittance along the various DOF. For instance, devising a beam transport system capable of supporting eigenemittances with large ratios has been proposed as a simple method to form flat beams $-$ beams with asymmetrical transverse emittances $-$ to suppress beamstrahlung at the interaction point of future electron-positron linear colliders~\cite{brinkmann-2001-a}. Likewise, the ability to control the partitioning of final emittances between the 3 DOFs could significantly improve the performances of compact high-gain free-electron lasers~\cite{cornacchia-2002-a}.

This Letter reports on the direct observation of the transfer of eigenemittances to projected emittances. Specifically, an electron beam with coupled beam dynamics between the transverse DOFs was prepared and accelerated to relativistic energies where it was then decoupled via a linear transformation. The resulting projected invariants were found to correspond to the two eigeninvariants associated with the initial coupled beam. This experiment validates the power of eigenemittances to prepare the initial beam distribution to attain the desired final emittance partition~\cite{yampolsky-2010-a}. Although previous experiments have demonstrated emittance repartitioning~\cite{piot-2006-a,groening-2014-a}, none have so far confirmed the direct mapping of eigenemittances to projected emittances.

The reported experiment pertains to the two transverse DOFs. We assume the transverse and longitudinal dynamics are uncoupled. Correspondingly, we describe the beam in four-dimensional (4-D) phase space with coordinates $\left[\mathbf{X}\equiv (x,x'\equiv p_x/mc),\mathbf{Y}\equiv (y,y'\equiv p_y/mc)\right]$ where $(x,p_x)$ and $(y,p_y)$ respectively refer to the horizontal and vertical position-momentum coordinates, and $m$ and $c$ are the electron rest mass and speed of light. The beam travels along the longitudinal coordinate $z$ so that the longitudinal momentum satisfies $p_z\gg(p_x,p_y)$. In the 4-D $\left(\mathbf{X},\mathbf{Y}\right)$ phase space the beam is statistically described by a $4\times4$ covariance matrix~\cite{ha2021bunch} 
\begin{equation}
\Sigma =\begin{bmatrix} \langle \mathbf{X} {\mathbf{X}}^T\rangle & \langle \mathbf{X} {\mathbf{Y}}^T\rangle \\  \langle \mathbf{Y} {\mathbf{X}}^T\rangle & \langle \mathbf{Y} {\mathbf{Y}}^T\rangle \end{bmatrix},   
\end{equation}
where $\langle \cdots \rangle$ indicates the statistical averaging over the phase-space PDF and the superscript $ \cdots^T$ stands for the transposition operator. When the transverse DOFs are decoupled, the $2\times 2$ off-diagonal blocks vanish ($\langle \mathbf{X} {\mathbf{Y}}^T\rangle=\langle \mathbf{Y} {\mathbf{X}}^T\rangle=0$) and the projected emittance, e.g., $\varepsilon_x\equiv \det(\langle \mathbf{X} {\mathbf{X}}^T\rangle)=[\langle{x^2}\rangle \langle{{x'}^2}\rangle-\langle{xx'}\rangle^2]^{1/2}$ is an invariant of motion in $(x,x')$ and similarly for $\varepsilon_y$ in $(y,y')$. In the most general case, the 4D emittance $\varepsilon_{ \mathrm {4d}}\equiv [\det(\Sigma)]^{1/2}$ is conserved under linear forces. More generally, owing to its positive-definite nature, $\Sigma$ can be diagonalized~\cite{williamson-1936-a} via a symplectic transformation $A$ as $A\Sigma A^T=\mbox{diag}(\beta^*_+\varepsilon_+, \varepsilon_+/\beta^*_+,\beta^*_-\varepsilon_-, \varepsilon_-/\beta^*_-)$ with $\varepsilon_{\pm}$ being the eigenemittances and $\beta^*_{\pm}>0$ are the betatron functions~\cite{courant-1958-a}. Specifically, the quantity $\sqrt{\beta_{\pm}\varepsilon_{\pm}/\gamma}$ (where $\gamma$ is the Lorentz factor) represents the electron-beam rms size at a waist. 
\begin{figure}[tttt!!!!!!!!!]
\centering
\includegraphics[width=0.95\columnwidth]{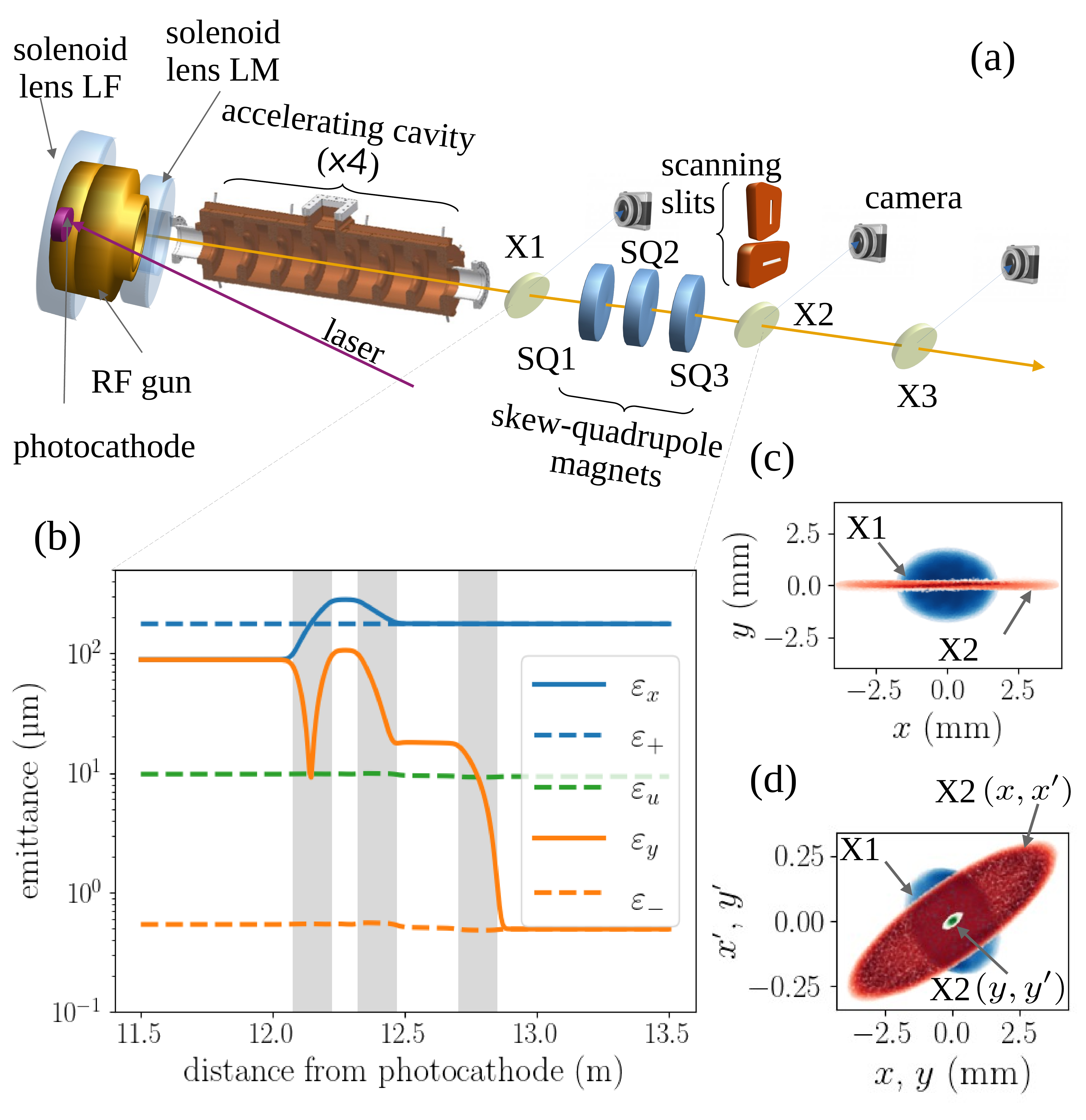}
\caption{\label{fig:beamline} Overview of the AWA beamline showing only components relevant to the present experiment (a) with simulated emittance evolution in the skew-quadrupole channel (b), and associated beam (c) and phase-space (d) distributions simulated at X1 and X2. In (a) the labels ``SQ$i$" and ``X$j$" respectively correspond to skew-quadrupole magnets and insertable scintillating screens. The diagnostics station located at X2 also comprises scanning slits for emittance measurements. In (d) the phase-space simulated at X1 is cylindrical-symmetric so that only $(x,x')$ distribution (blue) is reported while both $(x,x')$ (red) and $(y,y')$ are reported at X2. In (b) we introduce the uncorrelated emittance as $\varepsilon_u\equiv \sqrt{\varepsilon_{4d}}$  for convenience.}
\end{figure}
The eigenemittances are degenerate solutions of the characteristic equation of $\operatorname { det } \left( J_4 \Sigma- i \varepsilon _ { \pm } I \right) = 0$, where $i\equiv \sqrt{-1}$, $I$ is the $4\times 4$ identity matrix and $J_4 \equiv \left[ \begin{array} { c c } J & 0 \\ 0 & J \end{array} \right]$ with $J \equiv \left[ \begin{array} { c c } 0 & 1 \\ - 1 & 0 \end{array} \right]$. They are related to the projected emittances via~\cite{groening-2021-a}  
\begin{eqnarray}~\label{eq:eigen}
\varepsilon_{\pm} = \frac{1}{2}[\xi_+^{1/2} \pm \xi_-^{1/2}],
\end{eqnarray} 
where $\xi_{\pm}\equiv \varepsilon_x^2+\varepsilon_y^2+2\operatorname { det } (\langle \mathbf{X} {\mathbf{Y}}^T\rangle) \pm 2 \varepsilon _ {4d} $.  Equation~\ref{eq:eigen} gives a prescription to tailor the initial cross-plane correlation with given eigenemittance computed within the 4-D phase space distribution to ultimately $-$ after decoupling of the beam dynamics $-$ control the transverse projected emittance partition~\cite{carlsten-2011-a}. Such a capability is foreseen to have applications in the design of future high-energy particle accelerators~\cite{brinkmann-2001-a} and free-electron lasers~\cite{yampolsky-2010-a}. 
\begin{figure}[bb!]
\centering
\includegraphics[width=0.95\columnwidth]{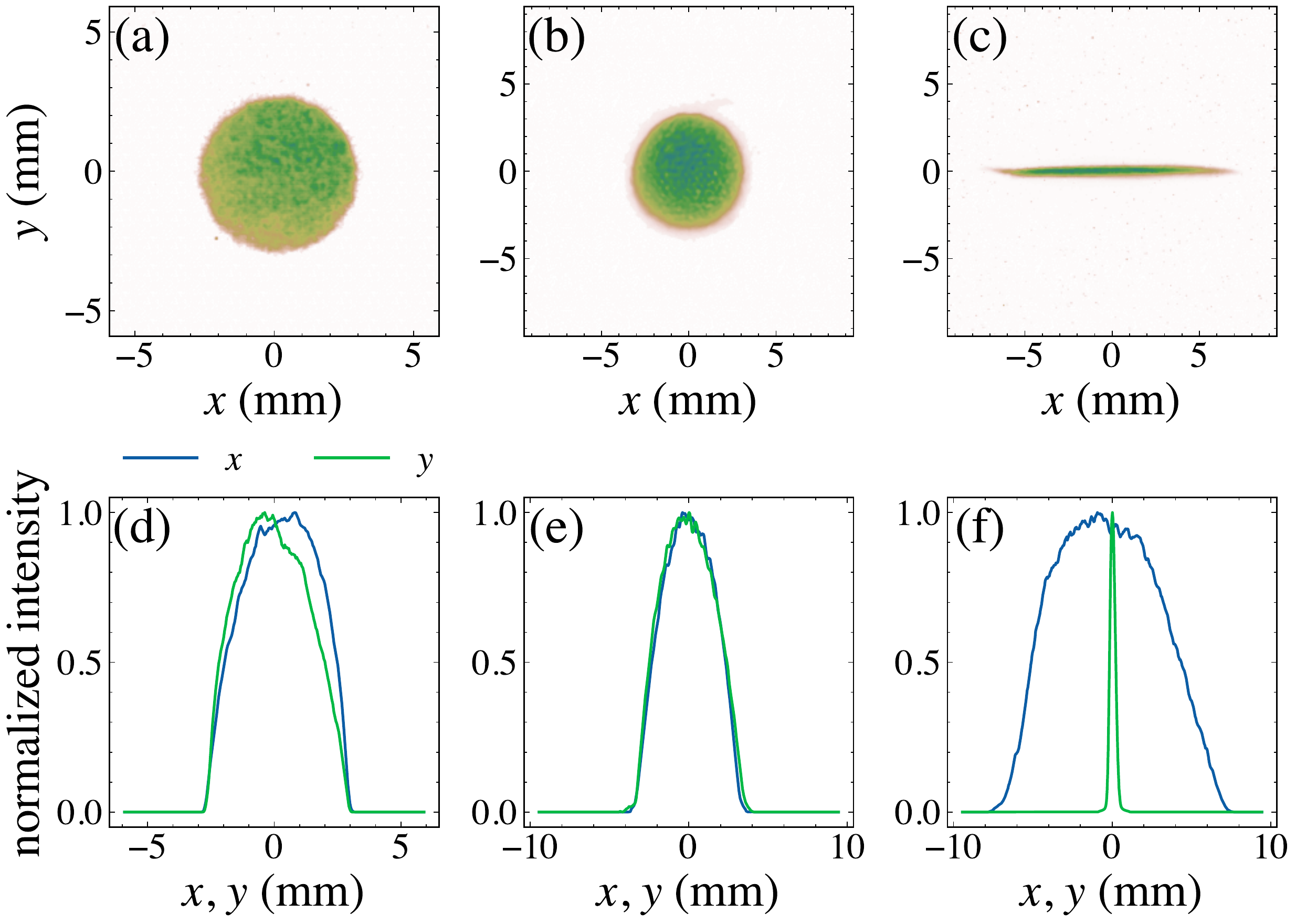}
\caption{\label{fig:measurement} Representative measurements of the laser distribution on the virtual cathode (a), and transverse electron-beam distribution recorded at X1 for the coupled beam (b) and at X2 for the decoupled beam (c). The lower row displays the horizontal (blue) and vertical (green) projection computed from the upper row. The solenoid lens ``LF" was set to produce $B_c=140$~mT and switching between coupled and uncoupled configuration is achieved by powering the skew quadrupole magnets (SQ1-3) to their nominal values.}\end{figure}

The emittance-transfer experiment was performed at the Argonne Wakefield Accelerator (AWA) diagrammed in Fig.\ref{fig:beamline}(a)~\cite{conde-2011-a}. The beamline consists of a radiofrequency (RF) photoinjector comprising a Caesium-Telluride (Cs$_2$Te) photocathode mounted in a 1+$\frac{1}{2}$ resonant cavity (the RF gun). In our experiment, $\sim 1$-nC electron bunches were emitted from the gun into a linear accelerator (linac) that boosts the beam energy to $\sim \kern-3pt {42}$~MeV. To produce a beam in an initial coupled state, the photocathode was immersed in an axial magnetic field $B_c\hat{\mathbf{z}} $ resulting in a particle emitted at position $\mathbf{r}^T\equiv(x,y)$ on the photocathode to acquire an ab-initio angular momentum, arising from the non-vanishing angular components of the vector potential $\mathbf A \simeq  \frac{B_c}{2}(-y\hat{\mathbf{x}}+x \hat{\mathbf{y}})$, given by $\mathbf{L}= \mathbf{r}\times e \mathbf{A} = \frac{eB_c}{2}r^2\hat{\mathbf{z}}$. Correspondingly, the $2\times2$ anti-diagonal blocks of the beam covariance matrix are given by $\langle \mathbf{X} {\mathbf{Y}}^T\rangle=-\langle \mathbf{Y} {\mathbf{X}}^T\rangle={\cal L} J$ where ${\cal L}\equiv \frac{\langle L \rangle }{2mc}$ is the beam magnetization, and the ensemble-averaged angular momentum is $\langle L\rangle= { e B _ { c } \sigma _ { c } ^ { 2 } }$ where $\sigma_c\equiv \sqrt{\mean{x^2}}=\sqrt{\mean{y^2}}$ is the transverse rms size of the axi-symmetric laser spot on the photocathode~\cite{burov-2002-a,kim-2003-a}. The ultraviolet laser pulse used to trigger the emission was transversely shaped to ensure the distribution of the emitted beam is cylindrical-symmetric and homogeneous~\cite{halavanau-2017-a}; the measured laser transverse distribution on the virtual photocathode $-$ a 1-to-1 optical image of the photocathode$-$ appears in Fig.~\ref{fig:measurement}(a). 
Likewise, the full-width half-maximum of the laser-pulse duration was $\sim \kern-3pt {400}$~fs to guarantee the evolution of the beam dynamics is dominated by the space-charge-driven expansion resulting in a quasi uniformly-charged ellipsoidal distribution with space-charge forces linearly dependent on the positions within the bunch~\cite{luiten-2004-a,musumeci-2008-a}. Such a ``blow-out" regime mitigates emittance growth arising from nonlinear space-charge effects. Downstream of the linac, three skew-quadrupole magnets SQ1-3 is used to apply a torque to cancel the incoming beam's angular momentum; see Fig.~\ref{fig:beamline}(b). In the process, the initially round beam is transformed into a flat beam with an asymmetric emittance partition~\cite{brinkmann-2001-a,kim-2003-a}; see Fig.~\ref{fig:beamline}(b-d). The numerical modeling of the beam-dynamics presented in Fig.~\ref{fig:beamline}(b-d), performed with the {\sc impact-t}~\cite{qiang-2006-a} program, predicts an ideal mapping of the eigenemittances into projected emittances [see Fig.~\ref{fig:beamline}(b)] allowed by the linear space-charge forces supported by the quasi-ellipsoidal distribution~\cite{kapchinskij-1959-a,lapostolle-1965-a}. Figure~\ref{fig:measurement}(b,c) compares the beam distributions recorded at X1 and X2 and illustrates the transformation of the incoming coupled cylindrical-symmetric beam in the ``flat" decoupled beam with a high transverse emittance (and aspect) ratio. 

\begin{figure}[t!]
\centering
\includegraphics[width=0.95\columnwidth]{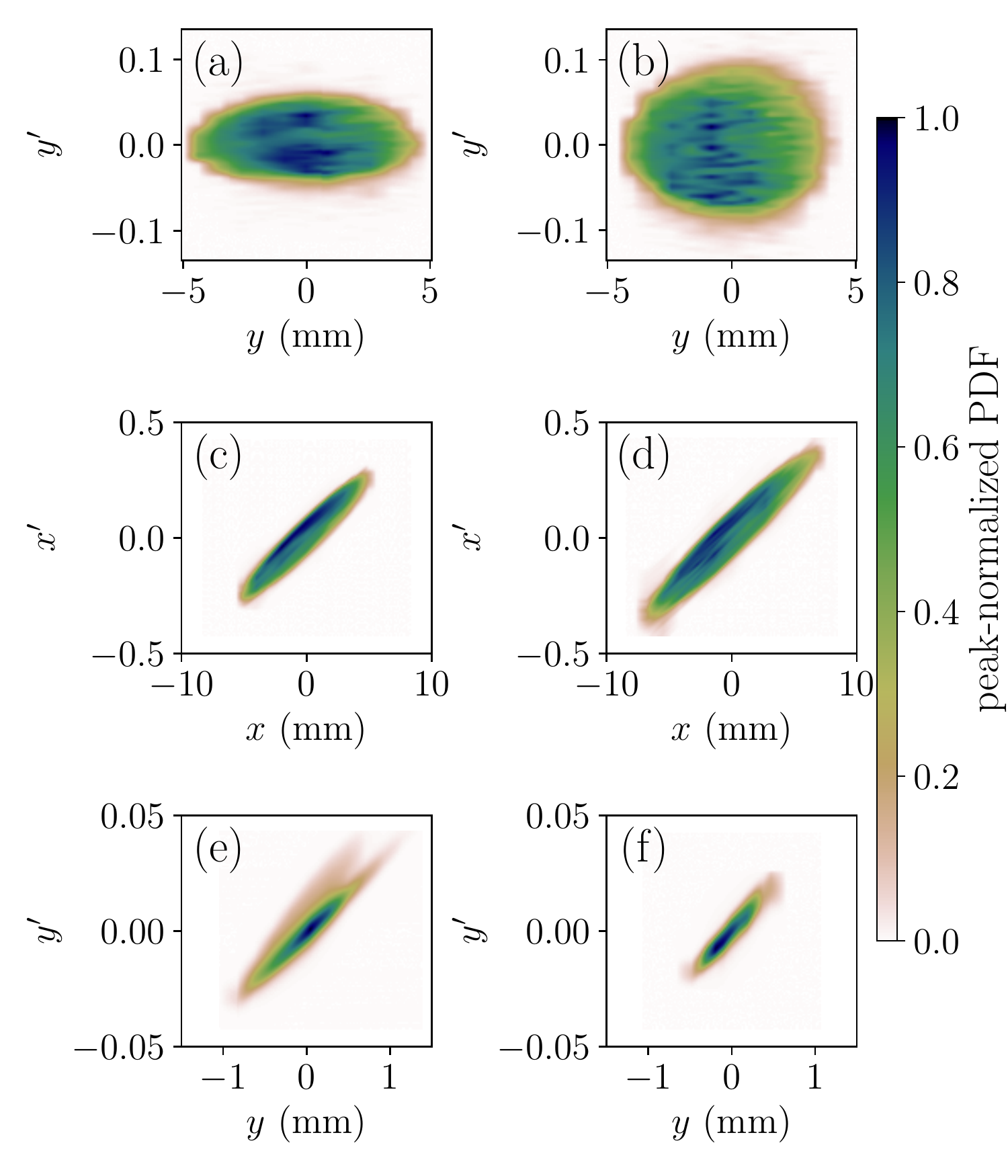}
\caption{\label{fig:phase_space} Measured phase-space PDFs of the electron bunch in coupled (a,b) and uncoupled states (c-f) for $B_c=79$~mT (a,c,e) and $B_c=140$~mT (b,d,f). Plots (a,b): $(y,y')$ PDF of the round beam; Plots (c,d) and (e,f) are respectively the $(x,x')$ and $(y,y')$ PDF of the final uncoupled flat beams.}
\end{figure}

Our experiment employs $B_c$ as a single knob to control the strength of the transverse coupling $\mathcal { L }$ to set the initial eigenemittance partition~\cite{groening-2014-a}. The photoemitted bunch (with charge $Q=1\pm 0.1$~nC)  is accelerated to $42\pm 0.5$~MeV before being decoupled by the the skew-quadrupole magnets SQ1-3 downstream of the linac. The post-linac beamline includes several diagnostics stations (X1-3). All 3 stations have 50-mm diameter Cerium-doped Yttrium aluminium garnet (Ce:YAG) electron imaging screens to measure the transverse distribution. The slit-scan method is used to measure the phase space. It uses horizontally- and vertically-scanning slits (with respective widths of 100 and 50~\textmu{m}) at X2 . These slits are scanned across the beam and the transmitted beamlets are recorded at X3 to provide a measurement of the transverse momentum spread associated with the beamlet at the transverse position of the slit. Therefore, the measurement can be used to reconstruct the transverse phase space at X2. In the case of an incoming coupled beam (SQ1-3 turned off) the transmitted beamlets shears, in addition to diverging, as it reach X3. The net shearing angle $\theta$ provides a measurement of  $\mathcal { L } =  \gamma \frac { \sigma _ {2} \sigma _ {3} \sin \theta } { D }$, where $\sigma _ {i}$ ($i=2,3$) stands for the rms beam sizes at X$i$, and $D$ is the distance separating X2 to X3~\cite{sun-2004-a}. The measured beamlet divergence at X2 associated with the sampled position across the beam at X1 can be used to reconstruct the phase space as illustrated in Fig.~\ref{fig:phase_space}(a,b) where the $(y,y')$ phase space of the incoming magnetized beam was measured for two values of magnetization. The phase spaces were reconstructed from 11 sampling positions of the slit and interpolated using linear splines. The $(x,x')$ phase-space is very similar due to the beam cylindrical symmetry beam and we report only the vertical plane as the smaller-width scanning slit provides a better resolution than the $(x,x')$ measurement. The $(y,y')$ phase-space PDFs shown in Fig.~\ref{fig:phase_space}(a,b) are uncorrelated, i.e. $\langle yy' \rangle \simeq 0$, which is expected as the upstream beamline was tuned to produce a waist close to the slit location for the coupled beam. The emittance directly computed from these ``projected" phase-space PDFs yields the projected emittance $\varepsilon_{ x}$.  Likewise, removing the averaged beamlet shear and analyzing the so-processed beamlets provide a direct measurement of the uncorrelated phase space PDF which provides $\varepsilon_{u}$~\cite{hannon-2019-a}. 

The decoupling of the beam dynamics is accomplished with three skew-quadrupole magnets having their strengths set to exert the net torque needed to remove the initial angular momentum. It is generally challenging to diagonalize the beam covariance matrix, and the decoupling transformation $M$ implemented by the skew-quadrupole magnets produces instead a block-diagonal matrix resulting in the final beam covariance matrix for the uncoupled state,
\[ \Sigma_f = M \Sigma \tilde { M } = \begin{bmatrix}  \varepsilon_+ T_+ & 0 \\ 0 & \varepsilon _- T_- \end{bmatrix} \mbox{~with~} T_{\pm} = \left[ \begin{array} { c c } \beta_{\pm}  & -\alpha_{\pm}  \\ - \alpha_{\pm}  & \gamma_{\pm}  \end{array} \right]\]
where $\beta_{\pm}>0$ are the betatron functions, $\alpha_{\pm}\equiv -\frac{1}{2} \frac{d\beta_{\pm}}{ds}$ measures the phase-space linear correlation and $\gamma_{\pm}\equiv (1+\alpha_{\pm}^2)/\beta_{\pm}$ so that $\det(T_{\pm})=1$ and the final projected emittance associated with $\Sigma_f$ are $(\varepsilon_{f,x},\varepsilon_{f,y})=(\varepsilon_+,\varepsilon_-)$ (note that the mapping can be swapped by changing the magnets polarity). The pair $(\alpha_\pm, \beta_\pm$) is also referred to as the Courant-Snyder (CS) parameters~\cite{courant-1958-a}. 
\begin{figure}[h!]
\centering
\includegraphics[width=0.95\columnwidth]{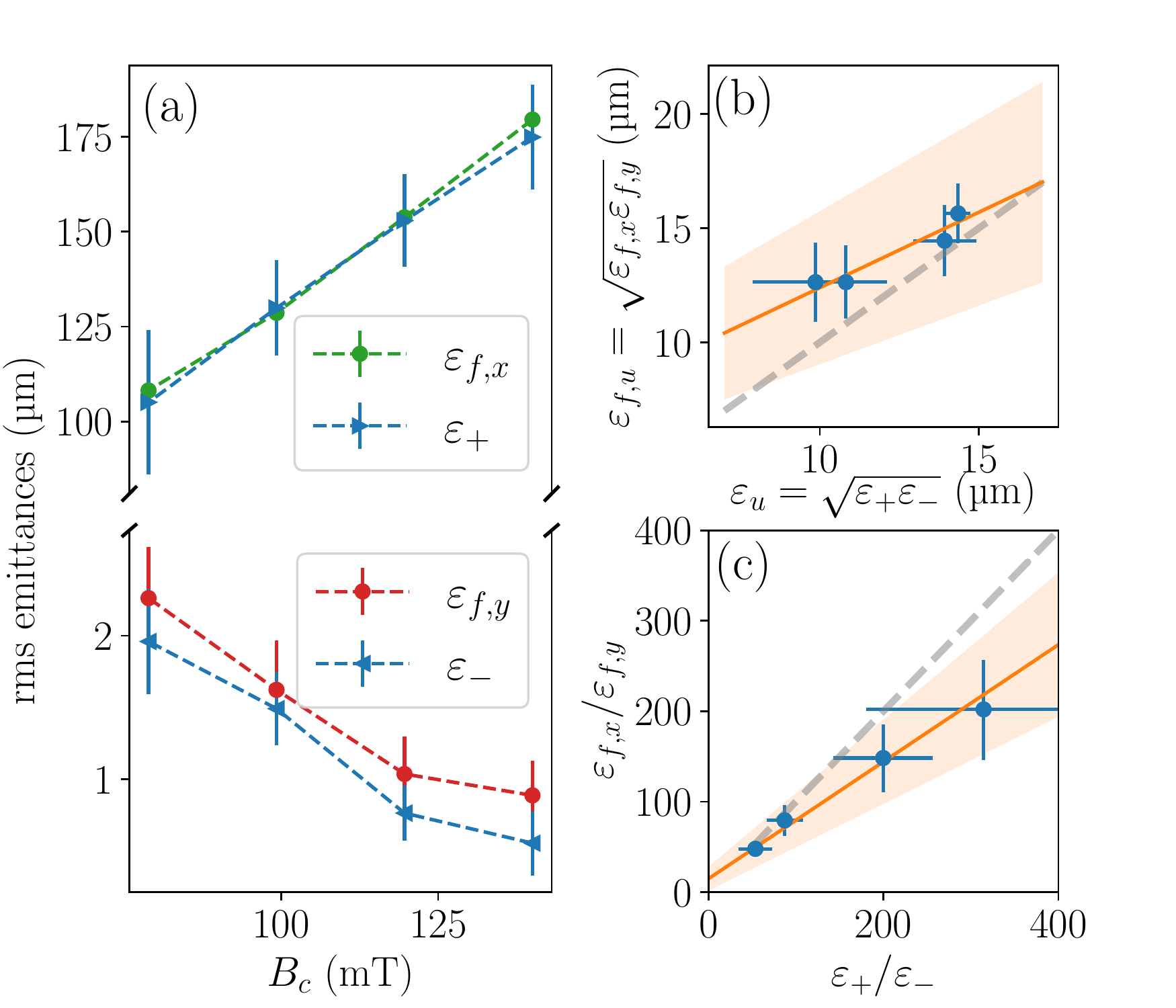}
\caption{\label{fig:mapping} Measurement of eigenemittances associated with the incoming coupled beam $(\varepsilon_+,\varepsilon_-)$ and the final projected emittances $(\varepsilon_x,\varepsilon_y)$ after decoupling  for different values of the magnetic field on the photocathode surface $B_c$ (a). Transfer of the uncorrelated emittances (b) and emittance ratio (c). The identity line is shown as a grey-dash line in (b) and (c) and the orange line with shaded area respectively correspond to a linear regression with associated $1\sigma$ confidence interval. }
\end{figure}
To devise the skew-quadrupole magnets settings, a beam-envelope fitting technique~\cite{mcdonald-1989-a}, using the screens  X1-3, was used to obtain the CS parameters at the entrance of the skew-quadrupole channel. These CS parameters were then used as inputs to optimize the skew settings to decouple the motion. The optimization was performed with {\sc impact-t} simulation considering the realistic quadrupole-magnet field profiles and taking into account space-charge effects. These settings of the skew strengths were further experimentally fine-tuned to ensure the beam remained "flat" when observed on X2 and X3 $-$ thereby experimentally confirming the coupling is fully canceled. Finally, slit-scan measurements were conducted to measure the horizontal and vertical phase space PDFs associated with the flat beam. Our choice to produce a vertically-flat beam stems from the higher-resolution diagnostics available for the vertical phase space. Figure~\ref{fig:phase_space}(c-f) displays such PDF for two cases of initial coupling. For large values of ${\cal L}$ the reconstructed phase spaces are highly unbalanced [the phase-space area of the $(x,x')$ is up to $\sim 200$ times larger than $(y,y')$)]; see Figure~\ref{fig:phase_space}(d,f). The asymmetry between the two final transverse phase spaces can be observed by examining Fig.~\ref{fig:phase_space}(c,d) and (e,f). 

From the measured coupling $\mathcal{L}$ and uncorrelated emittance $\varepsilon_{ \mathrm { u }}$, eigenemittances of the coupled beam are calculated from Eq.~\ref{eq:eigen} with $\xi\pm = 2(\varepsilon_{\mathrm{\perp}}^2 +{\cal L}^2 \pm \varepsilon_{4d})$ for a round beam and defining $\varepsilon_{\mathrm{\perp}}\equiv [\varepsilon_{ \mathrm { u }}^2 + {\cal L}^2]^{1/2},$ with  $\varepsilon_{u}\equiv\varepsilon_{4d}^{1/2}$, we obtain~\cite{kim-2003-a} $
\varepsilon_{\pm}=[\varepsilon_{ \mathrm { u }}^2 + {\cal L}^2]^{1/2}\pm {\cal L}. $
The measured eigenemittances are compared with the final projected emittances after the decoupling transformation $(\varepsilon_{f,x},\varepsilon_{f,y})$ in Fig.~\ref{fig:mapping}(a). We generally find an excellent agreement for the mapping $\varepsilon_+\mapsto \varepsilon_{f,x}$ and $\varepsilon_-\mapsto \varepsilon_{f,y}$. Figure~\ref{fig:mapping}(b) displays the evolution of the $\varepsilon_{ \mathrm { u }}$ obtained from the coupled ($\varepsilon_u=\sqrt{\varepsilon_-\varepsilon_+}$) and decoupled states $\varepsilon_{f,u}=\sqrt{\varepsilon_{f,x}\varepsilon_{f,y}}$. We note that the value of $\varepsilon_{ \mathrm { u }}$ differs for each case of $B_c$ as our experimental configuration does not provide an independent control over the cross-plane correlation imposed on the beam and the emittance-compensation process which minimizes the value of $\varepsilon_{ \mathrm { u }}$~\cite{carlsten-1993-a,serafini-1997-a,chang-2004,miginsky-2009-a}. Nevertheless the measured $\varepsilon_{u,f}$ and $\varepsilon_{u}$ agrees within $\lesssim 20$\% relative error.  Finally Fig.~\ref{fig:mapping}(d) shows that unprecedented transverse emittance ratios  $\varepsilon_{f,x}/\varepsilon_{f,y} \gtrsim 200$ $-$ were attained in our experiment. The mapping is near ideal with  some discrepancies observed for high values of $B_c$ due to the limited resolution of our diagnostics to resolve the small values of $\varepsilon_{f,y}$. \\

In summary, we have experimentally demonstrated the transfer of generalized kinematical invariants $-$ the eigenemittances $-$ associated with a coupled beam to invariants associated with the lower-dimension orthogonal degrees of freedom after decoupling the beam via a linear transformation. These results confirm the potential of these generalized invariants to precisely control the phase-space partition between the degrees of freedom via tailoring of initial cross-plane correlations. Our findings could be generalized to configurations with higher dimensionality (e.g. introducing coupling between all DOFs) as proposed  in~\cite{duffy-2016-a}. Our research could also have applications to Optics and Quantum Mechanics where similar invariants are often introduced to describe coupled-motion or mixed-states configurations.\\

This work was supported by the U.S. Department of Energy (DOE), Office of Science, under award No. DE-SC0018656 with Northern Illinois University and contract No. DEAC02-06CH11357 with Argonne National Laboratory (ANL). We also acknowledge support from the ``US-DOE-Japan cooperation in High-Energy Physics" program and computing resources provided on {\sc bebop},  a high-performance computing cluster operated by the Laboratory Computing Resource Center at ANL.




%
\bibliography{sample}

\end{document}